\documentclass{aa}
 \usepackage[utf8]{inputenc}
\usepackage{graphicx}
\usepackage{natbib}
\usepackage{enumerate}

\usepackage{float}
\makeatletter

\newcommand{\Rmnum}[1]{\expandafter\@slowromancap\romannumeral #1@}

\newcommand{\emm}[1]{\ensuremath{#1}}   
\newcommand{\emr}[1]{\emm{\mathrm{#1}}} 
\usepackage{aas_macros}

\newcommand{\HII}{H\mbox{\sc ~ii}}
\newcommand{\HI}{\emr{HI}}

\begin{document}
\title{A 500 pc filamentary gas wisp in the disk of the Milky Way}
 \author{Guang-Xing Li \inst{1}  \and Friedrich Wyrowski\inst{1} \and Karl
 Menten\inst{1} \and Arnaud Belloche\inst{1}} \institute{Max-Planck Institut
 f\"ur Radioastronomie, Auf dem H\"ugel, 69, 53121 Bonn, Germany}
\offprints{Guang-Xing Li, \email{gxli@mpifr-bonn.mpg.de}}

\abstract{Star formation occurs in molecular gas. In previous studies,
the structure of the molecular gas has been studied in terms of molecular
clouds, but has been overlooked beyond the cloud scale.
We present an observational study of the molecular gas at
$49.5^{\circ}<l<52.5^{\circ}$ and $-5.0 \;{\rm km\;s^{-1}}<v_{\rm
lsr}<17.4\;{\rm km\;s^{-1}}$. The molecular gas is found in the form of a huge
{\bf ($\gtrsim 500\;\rm pc$)} filamentary gas wisp. This has a large
physical extent and a velocity dispersion of $\sim 5 \;\rm km\;s^{-1}$. The eastern part of the
filamentary gas wisp is located $\sim 130\;\rm pc$ above the Galactic disk
(which corresponds to 1.5--4 e-folding scale-heights), and the total mass of the gas
wisp is {\bf $\gtrsim 1 \times 10^5 M_{\odot}$}.
It is composed of two molecular clouds and an expanding bubble. The
velocity structure of the gas wisp can be explained as a smooth quiescent
component disturbed by the expansion of a bubble. That the length of
the gas wisp exceeds by much the thickness of the molecular disk of
the Milky Way is consistent with the cloud-formation scenario in which the gas is
cold prior to the formation of molecular clouds.
Star formation in the filamentary gas wisp occurs at the edge of a bubble
(G52L nebula), { which is consistent with some models of triggered star
formation.}}
 
 
\titlerunning{500 pc Filamentary Gas Wisp}

\maketitle

\section{Introduction}
Molecular clouds belong to the densest and coldest parts of the Milky Way
interstellar medium \citep{1969ApJ...155L.149F,1977ApJ...218..148M}.
Shielded from interstellar radiation fields, they provide conditions necessary
for star formation to take place. Observationally, molecular clouds exhibit a
complicated, irregular, and filamentary morphology
\citep{1987ApJ...312L..45B,1979ApJS...41...87S,2000prpl.conf...97W,2008ApJ...680..428G,2010A&A...518L.103M},
and (sub)millimeter-line observations of molecular clouds suggest that the gas
in the clouds is moving supersonically.
Consensus has not been reached concerning the origin and nature of
molecular clouds.

It must be recognized that molecular gas is just one of the phases of the Milky
Way interstellar medium, and its evolution is determined by many processes that
occur in the disk. To understand it,
we must also look into the large-scale structure of the multi-phased
interstellar medium, and understand the cloud evolution within this context.

Both observational and theoretical approaches have been taken in this
direction. Observationally, the distribution of molecular gas in nearby galaxies
can be accessed through millimeter line mapping
\citep[e.g.][]{2013arXiv1304.1801S}, from which structures such as spiral arms,
filaments, and spurs can be identified. Theoretically, the structure of
the multi-phased interstellar medium in a galactic disk has been studied
through simulating the whole disk with different approaches
\citep{2002ApJ...570..132K,2006ApJ...647..997S,2011ApJ...730...11T,2011MNRAS.413.2935D,2013ApJ...764...36V},
complemented by analytical calculations \citep{2012ApJ...756...45L}. It was
found that filaments or spurs can be created through the combination of
gravitational instability of a galactic disk, galactic shear, and frequent
encounters/agglomeration between molecular clouds
\citep{2001MNRAS.327..663P,2013MNRAS.432..653D}.

In the Milky Way, studies of the structure of molecular gas have been confined
spatially to the molecular cloud scale or have been limited to the structure of
the spiral arms
\citep{1981MNRAS.194..809L,1987ApJ...319..730S,2009ApJS..182..131R,2009ApJ...699.1153R,2010ApJ...723..492R}.
This is partly due to the complicated morphology of molecular gas and partly
due to the superposition of the emission of molecular gas from different
structures along the line of sight. In spite of these difficulties, it is of
both observational and theoretical interest to identify large, coherent
molecular structures in the Milky Way apart from the spiral arms, since these structures are natural tracers of the large-scale gas circulation
in the Milky Way disk.
In contrast to the extragalactic case, where we are limited by the
resolution and sensitivity of the telescopes (and the filtering of
interferometers), for our Milky Way it is possible to study the kinematics of
the molecular gas and the associated star formation with in more detail.

The Milky Way interstellar medium has long been thought to be dynamic. Shells
and rims are generally found in the disks of the Milky Way and other galaxies
\citep{2006ApJ...649..759C}. It has been proposed that the expansion
of \HII$\,$ regions, which creates shells and rims, can collect the interstellar medium into
a gravitationally unstable state
\citep{1977ApJ...214..725E,1994A&A...290..421W,2002MNRAS.329..641W}, and trigger
star formation. The expansion of the bubbles can also energize the interstellar
medium of the Milky Way efficiently
\citep{1996ApJ...467..280N,2004RvMP...76..125M}.

In this work, we present an observational study of the region at
$49.5^{\circ}<l<52.5^{\circ}$ in the Milky Way. The molecular gas in the region
exhibits a high degree of coherence, and forms a filamentary gas wisp (gas
filament) with a length of $\sim 3^{\circ}$.
The eastern part of the filamentary gas wisp sits at the edge of a bubble and
is located at $\sim 0.75^{\circ}$ above the galactic plane. {This eastern part
is listed in the context of infrared bubbles as one of the ``favorites of the
Milky-Way-Project volunteers'' \citep{2012MNRAS.424.2442S}, and it was studied
in terms of the G52L nebula by \citet{2012ApJ...759...96B}, who claimed that it
may be the largest single \HII\ region in the Milky Way.} Based on several estimations 
\citep{2003ApJ...587..714W,2009ApJ...690..706A,2010ApJ...723..492R,2012ApJ...759...96B},
the filamentary gas wisp has a distance of $9.77\;\rm kpc$, which implies a
physical length of $\gtrsim 500$ pc. { This is $\sim 5$ times longer than the
Nessie Nebula reported by \citet{2010ApJ...719L.185J} }\footnote{Note
that \citet{2013AAS...22123401G} report a much larger length of ``many
hundreds of pc'' for the Nessie nebula. See also http://milkywaybones.org/
for more details.}.
The physical length of the filamentary gas wisp exceeds by much the
size of a molecular cloud, and this filamentary gas wisp is by far the largest coherent
molecular structure identified in the Milky Way. It
exhibits a coherent velocity structure, and is composed of several molecular
structures, including two molecular clouds and one expanding bubble structure.
We present observations and an analysis of the region (Sect. \ref{sec:obs},
\ref{sec:theregion}), followed by a detailed discussion focusing on the
implications on the dynamics of the Milky Way interstellar medium and the life
cycle of molecular gas (Sect. \ref{sec:discussion}). In Sect.
\ref{sec:conclusion} we conclude.

\section{Archival data}\label{sec:obs}

\begin{figure*}
\hspace{ 0.15 em}
\includegraphics[width=0.93 \textwidth]{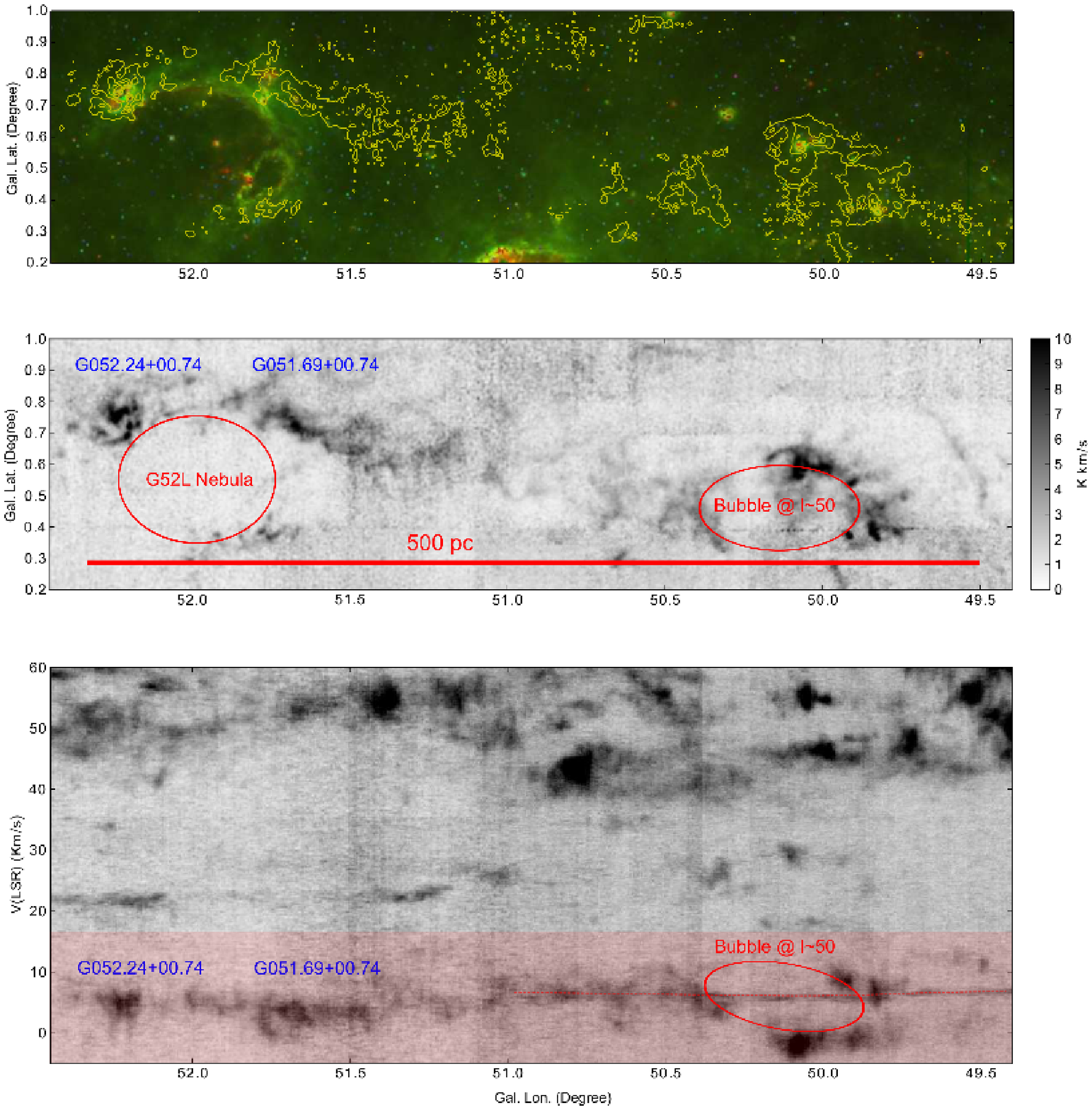}

\caption{\label{fig:region} {\bf Top panel:} 
Spitzer GLIMPSE \citep{2003PASP..115..953B} and MIPSGAL \citep{2009PASP..121...76C} three-color
image of the region.
Red: 24 $\mu$m, green: 8 $\mu$m, blue: 3.6 $\mu$m.
Overlaid contours are the velocity-integrated $^{13}$CO(1-0) emission
($-5.0\;{\rm km\;s^{-1}}<v_{\rm lsr}<17.4\;{\rm km\;s^{-1}}$) from the Galactic
Ring Survey \citep{2006ApJS..163..145J}.
Contours correspond to 3.5, 7.3, 11.2, 15 $\rm K\; km\;s^{-1}$. {\bf Middle
panel:} Velocity-integrated $^{13}$CO(1-0) map of the region integrated within
$-4.95 \;{\rm km\;s^{-1}}<v_{\rm lsr}<17.36\;{\rm km\;s^{-1}}$.  A scale bar of
500 pc is added assuming a kinematic distance of $9.8\;\rm kpc$. {\bf Bottom
panel:} Galactic-latitude-integrated $^{13}$CO(1-0) position-velocity map of the
region (integrated from $-0.2^{\circ}<b<1^{\circ}$).
The clouds G052.24$+$00.74, G051.69$+$00.74, the bubble at $l\sim 50^{\circ}$,
and the G52L nebula \citep{2012ApJ...759...96B} are indicated in the middle and bottom
panels.
The velocity range we used to produce the $^{13}$CO(1-0) integrated intensity
map is indicated in the bottom panel as the red shaded region.
The emission at $5.2\;{\rm km\;s^{-1}}<v_{\rm lsr}<7.2\;\rm km\; s^{-1}$ is due
to contamination from a different molecular cloud, and some of the emission lies
on the red dashed line.
This component has a smaller line width $\lesssim 0.5 \;\rm km\;s^{-1}$, which
implies that the contamination comes from a close-by cloud. This is
supported by its apparent diffuse morphology. Channel maps of the region are
provided in Appendix \ref{sec:app1}.}
\end{figure*}

We obtained 3.6 $\mu$m and 8 $\mu$m data from the
GLIMPSE project \citep{2003PASP..115..953B}, which is a fully sampled,
confusion-limited, four-band near-to-mid infrared survey of the inner Galactic
disk. We obtained 24 $\mu$m data from the MIPSGAL project
\citep{2009PASP..121...76C}, which is a survey of the Galactic disk with the
MIPS instrument on {\it Spitzer} at 24 $\mu$m and 70 $\mu$m. 

We obtained $^{13}$CO(1-0) molecular line data ($\nu_0=110.2\rm\; GHz$) from the
Galactic Ring Survey \citep{2006ApJS..163..145J}, which is a survey of the Milky
Way disk with the SEQUOIA multipixel array on the Five College Radio Astronomy
Observatory 14 m telescope, and covers a longitude range of $18
^{\circ}<l<55.7^{\circ}$ and a latitude range of |b|<$1^{\circ}$ with a spatial
resolution of $46\arcsec$.

\section{Results}\label{sec:theregion}
\subsection{ Region}
Figure \ref{fig:region} shows the {\it Spitzer} three-color image of the region
from $49.5^{\circ}<l<52^{\circ}$. The overlaid contours show the molecular gas
traced by $^{13}$CO(1-0). The CO emission in all the panels is integrated within
$-5.0 \;{\rm km\;s^{-1}}<v_{\rm lsr}<17.4\;{\rm km\;s^{-1}}$.
Several features can be identified. At $51.5^{\circ}<l<52.5^{\circ}$, there is a
bubble with a radius of $\sim 1^{\circ}$ \citep[G52L
nebula,][]{2012ApJ...759...96B}.
The molecular gas is situated to the north of the bubble. At
$51^{\circ}<l<52.5^{\circ}$, the molecular gas is organized in the form of two
molecular clouds (G052.24+00.74 and G051.69+00.74).
The cloud G052.24+00.74 has a roundish shape. This cloud is connected with
another molecular cloud, G051.69+00.74. This cloud has a more elongated
geometry, and star formation occurs only at its eastern part.

At $49.5^{\circ}<l<50^{\circ}$, there is noticeable contamination from gas
with $5.2 {\;\rm km\;s^{-1}}<v_{\rm lsr}<7.2 \;\rm km\;s^{-1}$ { (see the red
arrows in Fig. \ref{fig:channel}).
 The contaminating gas has an extremely narrow line width ($\lesssim
0.5\;\rm km\;s^{-1}$) }and tends to spread along the spatial direction. This
makes it easily distinguishable from the emission from the gas filament.
{This narrow line width implies that the emission comes from a close-by
cloud. This is supported by the fact that the contaminating gas has a more
diffuse morphology (see Appendix \ref{sec:app1} for $^{13}$CO(1-0) channel maps
of the region).} This distinction is similar to the supernova remnant
G016.05-0.57 studied in \citet{2011ApJ...741...14B}.

The two molecular clouds (G052.24$+$00.74 and G051.69$+$00.74) have a similar
velocity and velocity dispersion: the cloud G052.24$+$00.74 has $v_{\rm lsr}\sim
4.6\;\rm km\;s^{-1}$ and $ \delta v \sim 2.6 \;\rm km\;s^{-1}$ and the cloud
G051.69$+$00.74 has $v_{\rm lsr}\sim 3.6\;\rm km\;s^{-1}$ and $\delta v \sim
3.6 \;\rm km\;s^{-1}$ \citep[][]{2010ApJ...723..492R}.
In the position-position and position-velocity space, they are connected with
some wispy gas filaments (at $l\sim 52^{\circ}$ and $b\sim 0.8 ^{\circ}$ of the
top and middle panels of Fig.
\ref{fig:region}).
The similarity of the velocity dispersions and the proximity of the clouds in
position-velocity space imply that the two clouds are physically connected.

\begin{figure*}
\includegraphics[width=0.8 \textwidth]{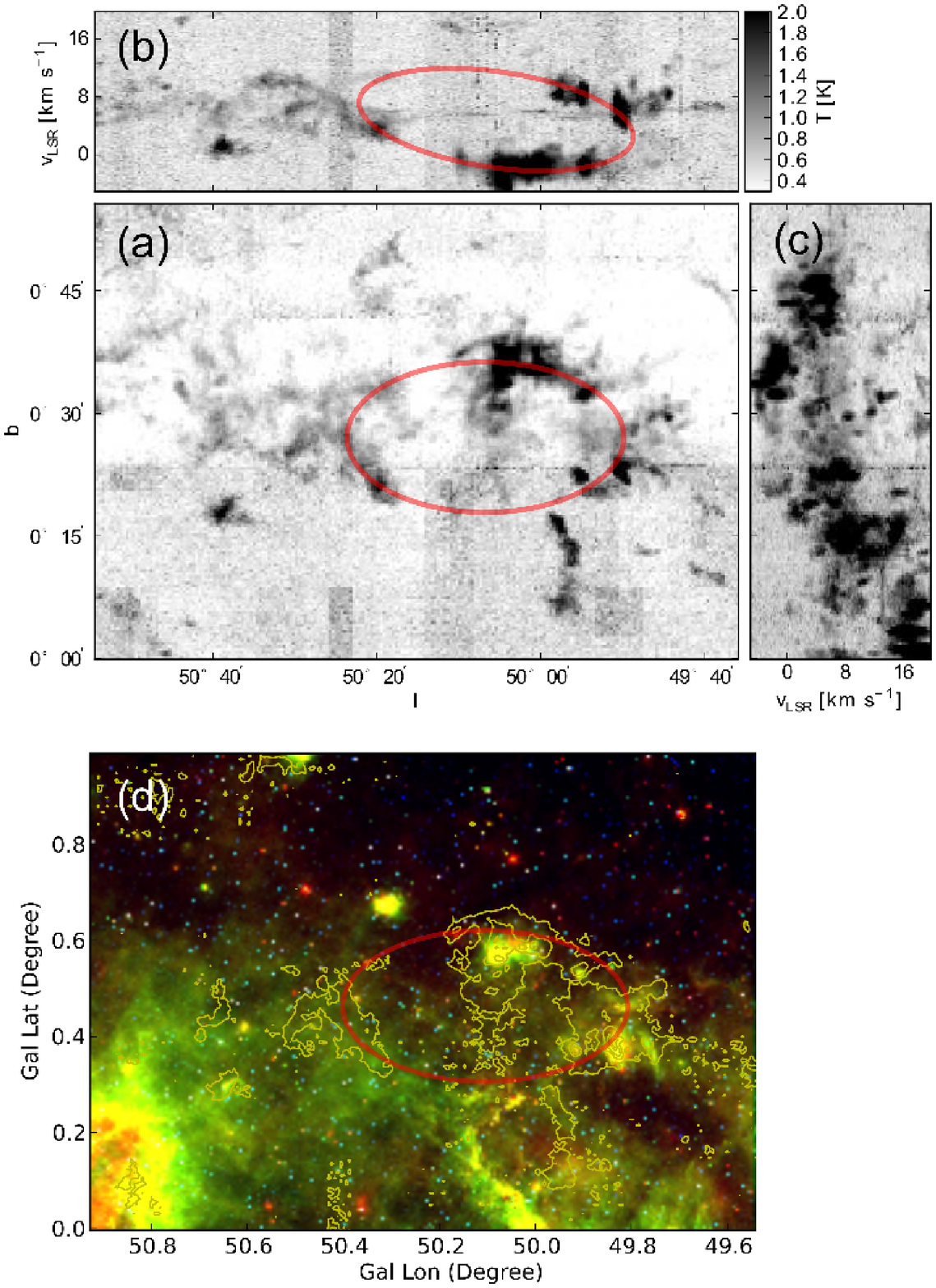}
\caption{ \label{figure:bubble} 
\textbf{(a)} Galactic longitude-latitude map of
the peak temperatures of the $^{13}$CO(1-0) data cube along the velocity axis.
\textbf{(b)} Galactic-longitude-velocity map of peak temperatures of
$^{13}$CO(1-0) along the galactic latitude axis.
\textbf{(c)} Velocity-galactic-latitude map of peak
temperatures of $^{13}$CO(1-0) along the galactic longitude
axis. 
\textbf{(d)} Spitzer GLIMPSE \citep{2003PASP..115..953B} and 
MIPSGAL \citep{2009PASP..121...76C} three-color image of the region.
Red: 24 $\mu$m, Green: 8 $\mu$m, Blue: 3.6 $\mu$m. 
Overlaid contours are the velocity-integrated $^{13}$CO(1-0) emission from the Galactic Ring
Survey \citep{2006ApJS..163..145J} ($-4.95
\;{\rm km\;s^{-1}}<v_{\rm lsr}<17.36\;{\rm km\;s^{-1}}$). Contours correspond to
3.5, 7.3, 11.2, 15 $\rm K\; km\;s^{-1}$. In \textbf{(a)} \textbf{(b)}, and
\textbf{(d)}, the bubble is indicated as a red ellipse.}
\end{figure*}

It can be readily seen from the $^{13}$CO(1-0) emission that this
double-cloud system belongs to a large filament (Fig. \ref{fig:region} {and
Appendix \ref{sec:app1}}).
The filamentary gas wisp is coherent in both the spatial and the velocity
direction, {  which makes it distinguishable from other molecular
structures, for instance the $\sim 50\;\rm km\;s^{-1}$ clouds (Appendix
\ref{sec:app2}).} Seen from the middle panel of Fig.
\ref{fig:region}, the filamentary gas wisp extends from $l=49.5^{\circ}$ to
$l=52.5^{\circ}$, which implies an angular extent of $\gtrsim 3^{\circ}$. Seen
from the bottom panel of Fig.
\ref{fig:region}, the filamentary gas wisp has a limited velocity range of $\sim
22 \;\rm km\;s^{-1}$ ($-5.0 \;{\rm km\;s^{-1}}<v_{\rm lsr}<17.4\;{\rm km\;s^{-1}}$).
Similar to the double-cloud system, all the molecular gas in the filamentary gas
wisp has a similar velocity dispersion. At $49.5^{\circ}<l<50.5^{\circ}$, the
filamentary gas wisp seems to be split in both the position-position and
position-velocity maps (middle and bottom panels of Fig. \ref{figure:bubble}). This coincides with
the presence of a bubble in the infrared band. To summarize, the filamentary gas
wisp is composed of two molecular clouds and one bubble.

Star formation occurs in different parts of the filamentary gas wisp. Star
formation in molecular clouds can be conveniently traced by 24 $\mu$m emission,
which originates from the dust heated by newly-born stars. In the {\it
Spitzer} image, this appears as red regions (Fig. \ref{fig:region} top and Fig.
\ref{fig:double}). In the cloud pair
G052.24$+$00.74 and G051.69$+$00.74, several star-forming sites can be
identified (Fig. \ref{fig:double}) based on the {\it Spitzer} 24 $\mu$m
emission, three of which are currently hosting compact \HII\ regions
\citep{1989ApJS...71..469L,2009A&A...501..539U}. { At
$51.5^{\circ}<l<52.5^{\circ}$, all the star-forming sites are located at the
edge of the G52L bubble.}

\subsection{Distance and size of the filament}
The distance to the region has been estimated by several authors.
Without trigonometric parallaxes, the distance to the region can be determined
with the kinematic method. One key step in determining the kinematic distance is
to resolve the kinematic distance ambiguity. 

There are different ways to resolve the ambiguity. Distance of the
filamentary gas wisp can be determined by studying the distance to
molecular clouds and \HII\ regions that belong to the filament. Using \HI\
self-absorption, \citet{2010ApJ...723..492R} found that the molecular clouds
G052.24$+$00.74 and G051.69$+$00.74 are located at the far distance. Using the
H$_2$CO absorption line, \citet{2003ApJ...587..714W} found that the \HII\ region
G52.23+0.74 is located at the far distance. Recently,
\citet{2009ApJ...690..706A} and \citet{2012ApJ...759...96B} studied the distance
to the \HII\ regions G052.201$+$0.752 and G052.259$+$0.700 with \HI\
emission/absorption method, and again found that they are at the far distance.
Therefore we conclude that the filamentary gas wisp is located at the far
distance, which is approximately 9.8 kpc. This suggests a galactocentric
distance of 8.2 kpc, and the filamentary gas wisp probably resides in
or around the Perseus arm.

Accordingly, the filamentary gas wisp we identified has a spatial extent of
$\sim 500$ pc. {If the filamentary gas wisp follows the spiral
structure, it is probably angled $\sim 45^{\circ}$ to our line of sight, and
therefore probably has a deprojected length a factor of $\sqrt{2}$ longer.
Therefore we conclude that the filamentary gas wisp has a length of $\gtrsim 500$ pc. }It is
one the of the largest coherent molecular structures in the Milky Way apart from the
spiral arms and the molecular ring.
The total mass of the filamentary gas wisp can be estimated using the
$^{13}$CO(1-0) emission. { To do this, we integrated over the
region with the line-of-sight integrated flux $I>3.5\rm\; K\;km\;s^{-1}$.
This corresponds to the first contour in the upper panel of Fig.
\ref{fig:region}. This mass estimate should be considered as a lower limit
since by selecting this threshold we only take the region with a high column
density ($N_{H_2}>1.75\times 10^{21}\;\rm cm^{-2}$) into account. Using Eq. 1--3
of \citet{2010ApJ...723..492R} and assuming an excitation temperature of
$10\;\rm K$, we obtain a total mass of $\sim 1\times 10^{5} M_{\odot}$ for the whole gas wisp
($49.5^{\circ}<l<52.5^{\circ}$).  According to \citet{2001ApJ...551..747S}, the
derived mass is only weakly sensitive to this choice of excitation temperature,
and in our case an excitation temperature of $20\;\rm K$ gives a mass of $\sim
0.6 \times 10^{5} M_{\odot}$.}

The two clouds at the eastern part of the filamentary gas wisp have $b\sim
0.74^{\circ}$. Using a kinematic distance of $9.77\;\rm kpc$, the double-cloud
system is $\sim 130\;\rm pc$ above the Galactic plane. { At a Galactocentric
distance of $\sim 8\;\rm kpc$, the molecular disk of the Milky Way has a FWHM
thickness of $90-180\;\rm pc$ \citep[at 7--8 kpc the FWHM is $\sim 90$ pc and at
8--9 kpc the FWHM is 186 pc,][]{2006PASJ...58..847N}. This corresponds to an
e-folding height of 38--80 pc. Therefore the height of the double-cloud system
is about 1.5--4 times  the e-folding height of the Galactic disk. }The
double-cloud system is a unique cloud system that is located far above the Galactic plane.
{ According to \citet{2012ApJ...759...96B}, one possible explanation is that
the material of the system has been displaced by the expansion of the G52L
nebula.}

\subsection{The bubble structure at $l\sim 50^{\circ}$}
Figure \ref{figure:bubble} shows the bubble structure at $l\sim 50^{\circ}$.
{ Its boundary is visible in both the 8 $\mu$m
emission, which traces polycyclic aromatic hydrocarbon
 (PAHs) and in the $^{13}$CO(1-0) emission.
The bubble is not easily visible at 24 $\mu$m, which traces hot dust heated
by a central star. Because of the apparent absence of the 24 $\mu$m emission,
the bubble structure does not seem to be driven by the expansion of a \HII\ region.
This is also supported by the absence of a diffuse \HII\ region in
the VGPS \citep{2006AJ....132.1158S} continuum image.}

It is more likely that the bubble structure is driven by the expansion of a
supernova.
The {\it Spitzer} image of the bubble resembles that of several supernova
remnants in the \citet{2006ApJ...649..759C} catalog. From panel (a) of Fig.
\ref{figure:bubble}, using the kinematic distance of 9.8 kpc, we estimate a
diameter of $1^{\circ}\sim180\;\rm pc$, and from panel (b) of Fig.
\ref{figure:bubble} we estimate a total expansion velocity of $10\;\rm km \;s
^{-1}$. These give an age of $50\;\rm Myr$. The energy of a possible supernova
explosion can be estimated through the Sedov-Taylor solution:
$E\sim r^3 v^2 \rho \sim 0.16 \times 10^{51}\;\rm erg $. The energy is
consistent with a supernova explosion. Here, a density of $10^{-24} \;
\rm g\; cm^{-3}$ is used, which is typical of warm neutral medium
\citep[cf.][]{2010A&A...520A..71B}.

\section{Discussion}\label{sec:discussion}
\subsection{Morphology of the filamentary gas wisp}
This giant molecular structure is among the largest molecular structures studied
in the Milky Way ($\gtrsim 500\;\rm pc$). The physical size of the gas
filamentary gas wisp is much larger than that of a typical molecular cloud
\citep[$\sim 10\;\rm pc$,][]{2010ApJ...723..492R}. The velocity dispersion along a single
line of sight in the filamentary gas wisp is not significantly different from
that of ordinary molecular clouds. 

The molecular gas in the filamentary gas wisp is concentrated
in the vertical (Galactic-latitude) direction and elongated along the horizontal
(Galactic-longitude) direction. { At different locations, the filamentary gas
wisp exhibits a different width. At $49.5^{\circ}<l<50.5^{\circ}$, the
filamentary gas wisp is split in the map, which makes it difficult to define its
width.
From the map, the cloud G052.24$+$00.74 appears to be more extended in the
vertical direction than the cloud G051.69$+$00.74. We therefore used its
vertical extent as an estimate of the width of the filamentary gas wisp. 
The vertical extent of the cloud G051.69$+$00.74 is measured for the region with
$I> 3.5\;\rm K\;km\;s^{-1}$ ($N_{\rm H_2}>1.75\times 10^{21}\;\rm cm^{-2}$).
This corresponds to the first contour in the upper panel of Fig.
\ref{fig:region}. We found that the cloud extends from $b\sim 0.65^{\circ}$
to $b\sim 0.82^{\circ}$. From this we estimated a diameter of 30 pc, which
implies an aspect ratio of $\sim 600/30=20$ for the gas wisp.} The gas wisp is one of the most
 elongated molecular structures found in the Milky Way \citep[see also
 the Nessie nebula,][]{2010ApJ...719L.185J}.
{ The width of the filamentary gas wisp is narrower than the FWHM
thickness of the molecular disk of the Milky Way, which is about 90--180 $\;\rm pc$
\citep[][]{2006PASJ...58..847N} at a Galactocentric distance of $\sim 8$ kpc.}

Similar large-scale
molecular structures have been observed in other galaxies. In spiral
galaxies, elongated gas condensations are frequently observed. They can be
seen as narrow dark lanes that extend perpendicular to the spiral arms
\citep{1970IAUS...38...26L,1970IAUS...39...22W}. The exact definitions of
dust lanes or spurs differ in the literature. However, in most cases spurs refer
to the objects whose widths are similar to that of spiral
arms \citep{1980ApJ...242..528E}. In our case, the filamentary gas wisp should
not be termed a spur because its width is narrower than the width of
the spiral arms of a typical galaxy, which is $\sim 500$ pc \citep{2011ApJ...726...85E}. 

In our case, the filamentary gas wisp is about one or two
orders of magnitudes longer than ordinary molecular clouds, but is
still much narrower than the spurs in galaxies. Therefore we propose that
the filamentary gas wisp is a new object that is yet to be classified. Because
of this, we termed it a gas wisp in this work to emphasize its elongated
morphology.

Even though the thickness of the
filamentary gas wisp is similar to the resolution of the PAWS
survey \citep{2013arXiv1304.1801S} of M51,
filamentary gas wisps of this size would not be detected. This is because the
survey is only sensitive to objects with a mass $\gtrsim 1.2 \times 10^5 M_{\odot}$ and the
clouds in the filamentary gas wisp are only $\sim 10^{4} M_{\odot}$. However,
similar large-scale gas structures in nearby galaxies are probably suitable
targets for ALMA thanks to its improved sensitivity.
Nearby face-on galaxies are expected to be excellent sites for studying these
gas wisps since line-of-sight confusion can be avoided. A future project at ALMA
targeting at the molecular gas in nearby face-on galaxies is expected to resolve
similar gas
condensations and provide a more
 complete picture of the structure of molecular gas in galaxies.

\subsection{Implications for the formation of molecular clouds}
The formation and evolution of molecular clouds is one of the fundamental
problems in interstellar medium studies. To account for the short formation
timescale of molecular clouds, two scenarios have been proposed. The first
scenario involves colliding flows.
In this scenario, molecular clouds form from diffuse gas (warm neutral medium)
collected into a dense phase (cold neutral medium/molecular medium) through
colliding flows
\citep{2005A&A...433....1A,2006ApJ...648.1052H,2007ApJ...657..870V,2010ApJ...715.1302V,2012ApJ...759...35I}.
The molecular gas can form quickly in the converging flows because of
dynamically-triggered thermal instability.

The second scenario has been proposed by
\citet{2001MNRAS.327..663P} and \citet{2013MNRAS.432..653D}. In this scenario,
the gas is already relatively dense and cold prior to becoming a giant molecular cloud.
According to \citet{2001MNRAS.327..663P}, there is expected to be copious cold
gas in the inter-arm regions since the
circulation of the molecular gas is a process that occurs at the disk scale. To
describe the global circulation of the gas, we divided it into the in-arm phase
in which the gas is situated inside the spiral arms, and the inter-arm phase in
which the gas is situated in the inter-arm regions.
As discussed in \citet{2001MNRAS.327..663P}, in the inter-arm phase, the cold
gas exists in the form of wisps, and the gas in these wisps will show up as
giant molecular clouds during the spiral-arm phase. This has been largely
confirmed by the simulations of \citet{2013MNRAS.432..653D}, which track the
evolution of single molecular clouds. These authors found that molecular
clouds begin to disperse as they leave the spiral arm.
Due to differential shear, the molecular clouds are transformed into filamentary
gas wisps in the inter-arm region.
Since the shear occurs at a large scale, we expect to see gas wisps whose
physical scale exceeds the thickness of the Milky Way disk. In our case, the
physical length of the filamentary gas wisp ($\gtrsim 500$ pc) is much larger
than the scale-height of the Milky Way molecular disk. This is consistent with
the cloud-formation scenario by
\citet{2001MNRAS.327..663P} and \citet{2013MNRAS.432..653D}.
Such large-scale structures are also observed in other numerical simulations of
galactic disks
\citep{2000ApJ...536..173T,2002ApJ...570..132K,2006MNRAS.367..873D,2006MNRAS.371.1663D,2006ApJ...647..997S,2009ApJ...700..358T,
2012MNRAS.420.3490C}.

On the other hand, it is difficult to understand the filamentary gas wisp in the
converging flow scenario. In this scenario, molecular gas forms from the
converging \HI\ gas through the dynamically-triggered thermal instability.
As summarized in \citet{2012MNRAS.425.2157D}, the sources of the converging
flows can be stellar winds or supernovae
\citep{2000ApJ...532..980K,2008ApJ...689..290H,2011arXiv1111.1859N}, turbulence
in the interstellar medium \citep{1999ApJ...527..285B}, spiral shocks
\citep{1982ApJ...259..133L}, and gravitational instability. In our case, the
filamentary gas wisp cannot be created by converging stellar winds, supernovae,
or turbulence in the interstellar medium, since these mechanisms are local and
cannot create structures that are larger than the thickness of the Milky Way disk. { Spiral
shocks and
gravitational instability might create conditions favorable for converging
flows to occur.}
However, to access these possibilities we need to simulate converging flows in a
galactic context and properly quantify the role of the dynamically-triggered
thermal instability in the formation of molecular gas.
This task has not been achieved yet. { Either the converging flow scenario
is unable to explain how filamentary gas wisps form, or our current
understanding of converging flows in a galactic context is incomplete.}

\subsection{Star formation in the molecular cloud pair G0524.2$+$00.74 and
G051.69$+$00.74} 
It is unclear to what extent molecular clouds are gravitationally bound.
Gravity is important at a variety of physical scales during star formation.
According to \citet{2010ApJ...723..492R}, the clouds
G052.24$+$00.74 and G051.69$+$00.74 have virial parameters of $0.29$ and $0.33$,
respectively. This means that both clouds are gravitationally bound
\footnote{Here and in  \citet{2010ApJ...723..492R} the virial parameter $\alpha$
of a molecular cloud is the ratio of its virial mass $M_{\rm vir}$ to its mass.}.

\begin{figure*}
\includegraphics[width=0.9 \textwidth]{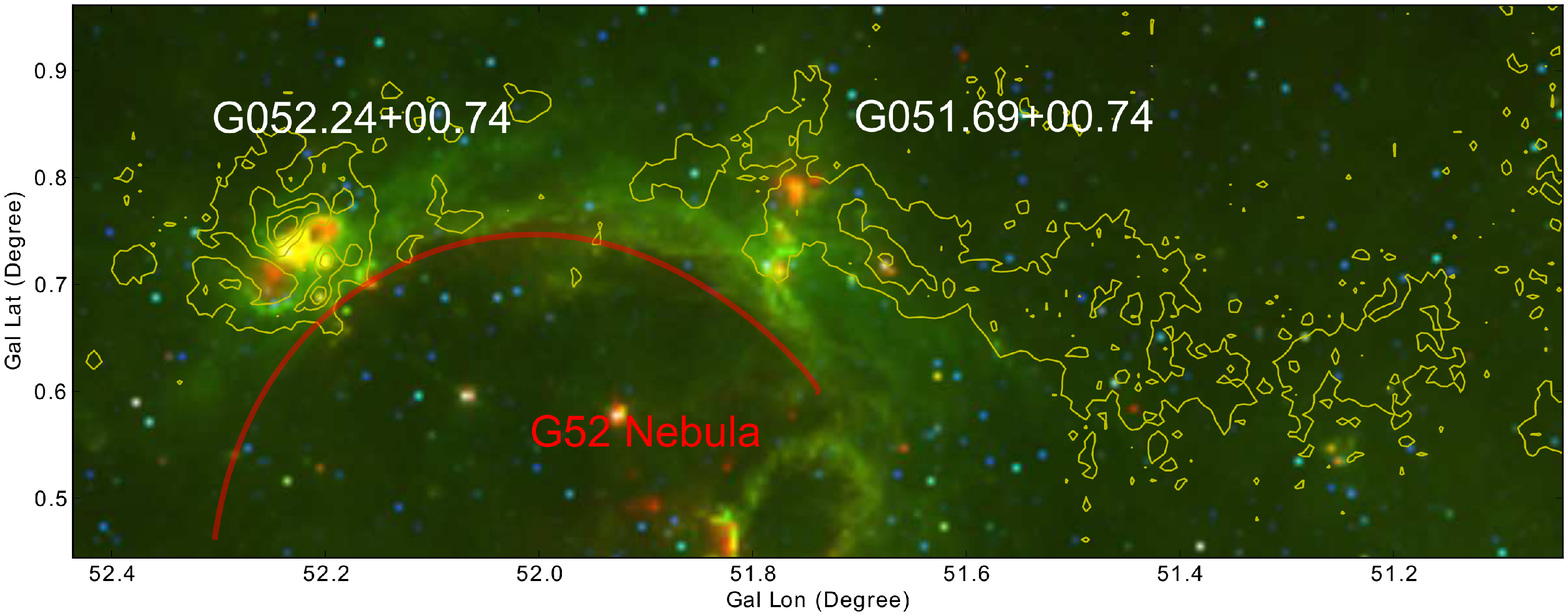}
\caption{Spitzer GLIMPSE \citep{2003PASP..115..953B} and 
MIPSGAL three-color image of the clouds G052.24+00.74 and G51.69+00.74.
Red: 24 $\mu$m, green: 8 $\mu$m, blue: 3.6 $\mu$m.
Overlaid contours are the velocity-integrated $^{13}$CO(1-0) emission ($-4.95
\;{\rm km\;s^{-1}}<v_{\rm lsr}<17.36\;{\rm km\;s^{-1}}$) from the Galactic Ring
Survey \citep{2006ApJS..163..145J}.
Contours correspond to 3.5, 7.3, 11.2, 15 $\rm K\; km\;s^{-1}$. The cloud
G052.24+00.74 and G51.69+00.74 as well as the G52L nebula are indicated. 
\label{fig:double}}
\end{figure*}

Star formation takes place in several sites, which is traced by the 24 $\mu
m$ emission in the {\it Spitzer} image. As indicated in  Fig. \ref{fig:double}, all
these sites seem to be located at the edge of a bubble (G52L nebula).

The connection between the location of the star-forming sites and the edge of
the bubble agrees with the statistical study of
\citet{2012MNRAS.421..408T}, in which a significant overdensity of young stellar
objects toward the edges of the bubbles was found.
This is consistent with the collect-and-collapse scenario of triggered star
formation \citep{1977ApJ...214..725E, 1994A&A...290..421W}.

\section{Conclusions}\label{sec:conclusion}
We studied a giant coherent molecular structure (a filamentary gas
wisp) at $49.5^{\circ}<l<52.5^{\circ}$. The eastern part of the
filamentary gas wisp is located $\sim 130\;\rm pc$ above the Galactic disk
(which corresponds to 1.5--4 e-folding scale-heights), and the total mass of the gas
wisp is $\gtrsim 1 \times 10^5 M_{\odot}$. Apart from the spiral arms and the
molecular ring, this is among the largest coherent molecular structures identified in the Milky
Way. The velocity structure of the filamentary gas wisp is coherent and smooth
at $50.5^{\circ}<l<52.5^{\circ}$, and at $49.5^{\circ}<l<50.5^{\circ}$, the
gas wisp is disturbed by a bubble structure. This
might be caused by a supernova. The overall velocity structure of
the filamentary gas wisp can be understood as a quiescent filamentary gas wisp
disturbed by the expansion of a bubble. The eastern part of the filamentary gas
wisp is composed of a system of two molecular clouds (G052.24$+$00.74 and
G051.69$+$00.74) and is located  $\sim 130\;\rm pc$ above the Galactic plane.

Star formation already takes place in several parts of this filamentary gas
wisp.
In the cloud pair G052.24$+$00.74 and G051.69$+$00.74, nearly the entire star
formation occurs at the edge of a bubble \citep[G52L
nebula,][]{2012ApJ...759...96B}. This is consistent with the
collect-and-collapse scenario of triggered star formation
\citep{1977ApJ...214..725E,1994A&A...290..421W}, and can be understood in the
statistical context of \citet{2012MNRAS.421..408T}.

The discovery of this filamentary gas wisp, whose length exceeds the thickness
of the molecular disk of the Milky Way, suggests that the formation and evolution of molecular clouds is a phenomenon that occurs at the
disk scale.
The large physical extent is consistent with the cloud-formation scenario by
\citet{2001MNRAS.327..663P} and \citet{2013MNRAS.432..653D}, in which the gas
that constitutes the molecular clouds is already relatively cold prior to the cloud
formation.

We are currently unable to answer how representative this filamentary gas wisp
is in the Milky Way disk. One reason is that we are restricted by the line-of-sight
confusion, and the filamentary gas wisp is fragile in nature. In our case, at
$49.5^{\circ}<l<50.5^{\circ}$, the filamentary gas wisp is already being
destroyed by the expansion of a bubble structure.
 It is possible that a significant fraction of the gas in the Milky Way exists
 in this form during at least part of its lifetime.
{ Another difficulty is to properly quantify the coherence of molecular
structures beyond the cloud scale. A position-velocity plot of the
$^{13}$CO(1-0) data from the same region shown in the bottom panel of Fig. \ref{fig:region} reveals
filamentary structures at $v_{\rm lsr}\sim 50\;\rm km\;s^{-1}$. { From a
visual inspection we found that these structures are not as coherent as the
filamentary gas wisp (see Appendix \ref{sec:app2} for a comparison).} In
general one cannot yet quantify the coherence of
molecular structures in the Milky Way.
More studies of the morphology of the molecular gas in both the Milky Way and
other galaxies with improved observations and analyses are needed to fully
understand the circulation of molecular gas at large scales.
}

\begin{acknowledgements} 
Guang-Xing Li is supported for this research through a stipend from the
International Max Planck Research School (IMPRS) for Astronomy and Astrophysics
at the Universities of Bonn and Cologne.

This publication makes use of molecular line data from the Boston
University-FCRAO Galactic Ring Survey (GRS). The GRS is a joint project of
Boston University and Five College Radio Astronomy Observatory, funded by the
National Science Foundation under grants AST-9800334, AST-0098562, \&
AST-0100793. This work is based in part on observations made with the Spitzer
Space Telescope, which is operated by the Jet Propulsion Laboratory, California
Institute of Technology under a contract with NASA.

We thank James Urquhart and Malcolm Walmsley for careful readings of our paper
and for many insightful comments. We thank the referee Adam Ginsburg for several
thorough and careful reviews of the paper and for his insightful comments. 
\end{acknowledgements}

\begin{appendix}
\section{Channel map of the filamentary gas wisp} \label{sec:app1}
\begin{figure*}
\includegraphics[width=0.95 \textwidth]{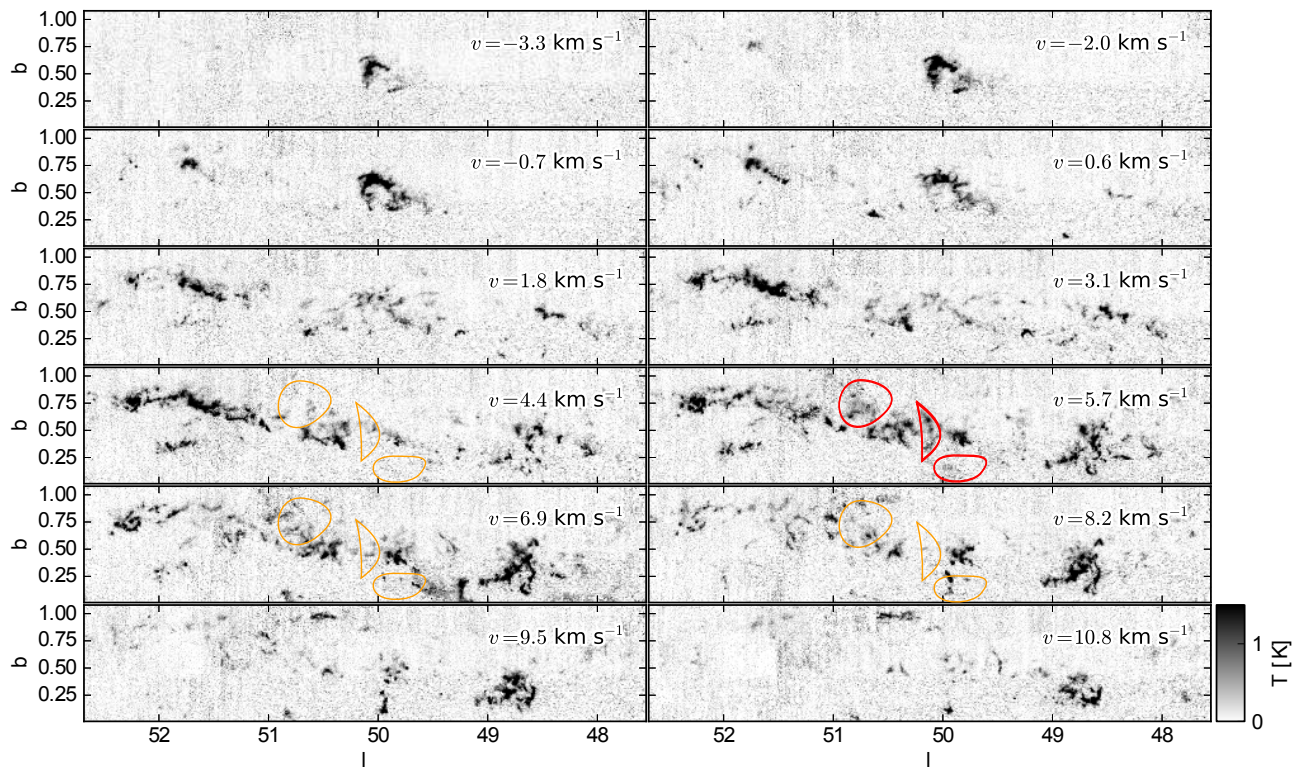}
\caption{{ Channel maps of the GRS \citep{2006ApJS..163..145J} $^{13}$CO(1-0)
emission of the region. The contaminating gas has a narrow ($\sim 0.5\;\rm
km\;s^{-1} $) line width, and is detected only in single channels in the map.
Some contamination from local clouds is indicated with red circles as
examples, {and these red circles are also plotted in yellow in the adjacent
velocity channels for comparison.}  }
\label{fig:channel}}
\end{figure*}

{ In Fig. \ref{fig:channel} we present the channel maps of the $^{13}$CO(1-0)
emission from the GRS \citep{2006ApJS..163..145J} survey. Some
contamination from local clouds is indicated. 
}

\section{A comparison with CO emission from the $\sim 50\;\rm
km\;s^{-1}$ component}\label{sec:app2}
{
To demonstrate the coherence of our filamentary gas wisp, in Fig.
\ref{fig:60kms} we present a map of the same region with the velocity integrated
within  $29.5 \;{\rm km\;s^{-1}}<v_{\rm lsr}<73.3\;{\rm km\;s^{-1}}$. Seen from
the $^{13}$CO(1-0) emission, this component is composed of individual
patches of molecular clouds and is not as coherent as our filamentary gas wisp.
The distance to the region is $\sim 5.3$ kpc. 

\begin{figure*}
\includegraphics[width=0.95 \textwidth]{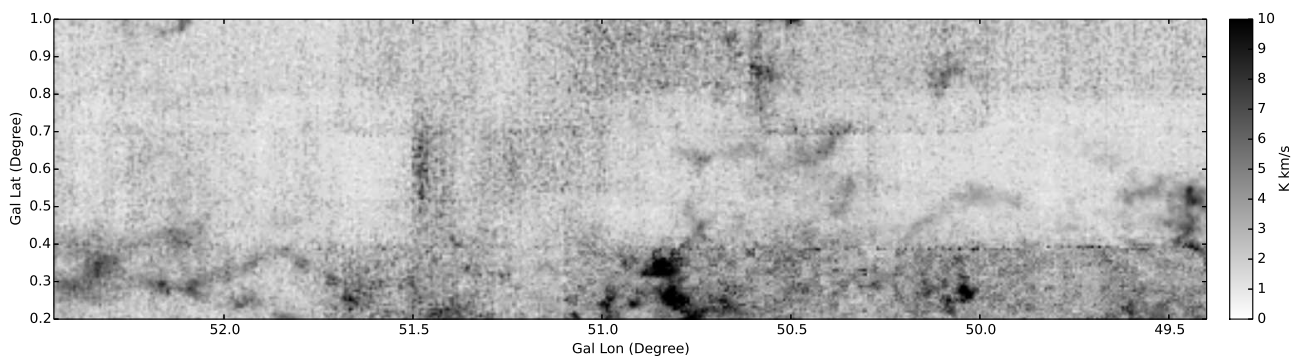}
\caption{{Velocity-integrated $^{13}$CO(1-0) map of the $\sim 50\;\rm
km\;s^{-1}$ clouds.
The emission is from the same region as shown in Fig. \ref{fig:region} and is integrated
within $29.5 \;{\rm km\;s^{-1}}<v_{\rm lsr}<73.3\;{\rm km\;s^{-1}}$.
\label{fig:60kms}}}
\end{figure*}
}

\end{appendix}

\end{document}